\documentclass[runningheads]{llncs}
\usepackage{graphicx}
\usepackage[english]{babel}
\usepackage[utf8]{inputenc}
\usepackage{booktabs}
\usepackage{makecell} 
\usepackage{arydshln}
\usepackage{amsmath}
\usepackage[caption=false,font=footnotesize]{subfig}
\usepackage[export]{adjustbox}
\usepackage{xcolor}
\usepackage[misc]{ifsym}

\usepackage{color}

\usepackage{graphicx}
\usepackage{hyperref}

\usepackage[inline]{enumitem}

\newcommand{\SKIP}[1]{}

\begin{document}
\title{A Hierarchical Multi-Task Approach \\ to Gastrointestinal Image Analysis}
\titlerunning{A Hierarchical MT Approach to Gastrointestinal Image Analysis}

\author{Adrian Galdran\inst{1} \and
Gustavo Carneiro\inst{2} \and
Miguel A. Gonz\'alez Ballester\inst{3,4}}
\authorrunning{A. Galdran et al.}
\institute{Dpt. of Computing and Informatics, Bournemouth University, UK, \email{agaldran@bournemouth.ac.uk} \and
Australian Institute for Machine Learning, University of Adelaide, Australia \and
BCN Medtech, Dept. of Information and Communication Technologies, \\Universitat Pompeu Fabra, Barcelona, Spain \and
ICREA, Barcelona, Spain}

\maketitle              %
\begin{abstract}
A large number of different lesions and pathologies can affect the human digestive system, resulting in life-threatening situations. 
Early detection plays a relevant role in the successful treatment and the increase of current survival rates to, \textit{e.g.}, colorectal cancer. 
The standard procedure enabling detection, endoscopic video analysis, generates large quantities of visual data that need to be carefully analyzed by an specialist. 
Due to the wide range of color, shape, and general visual appearance of pathologies, as well as highly varying image quality, such process is greatly dependent on the human operator experience and skill. 
In this work, we detail our solution to the task of multi-category classification of images from the gastrointestinal (GI) human tract within the 2020 Endotect Challenge. 
Our approach is based on a Convolutional Neural Network minimizing a hierarchical error function that takes into account not only the finding category, but also its location within the GI tract (lower/upper tract), and the type of finding (pathological finding/therapeutic intervention/anatomical landmark/mucosal views' quality).
We also describe in this paper our solution for the challenge task of polyp segmentation in colonoscopies, which was addressed with a pretrained double encoder-decoder network. 
Our internal cross-validation results show an average performance of 91.25 Mathews Correlation Coefficient (MCC) and 91.82 Micro-F1 score for the classification task, and a 92.30 F1 score for the polyp segmentation task. 
The organization provided feedback on the performance in a hidden test set for both tasks, which resulted in 85.61 MCC and 86.96 F1 score for classification, and 91.97 F1 score for polyp segmentation. At the time of writing no public ranking for this challenge had been released.
\end{abstract}

\begin{figure*}[t]
\centering
\subfloat[]{\includegraphics[width = 0.24\textwidth,valign=c]{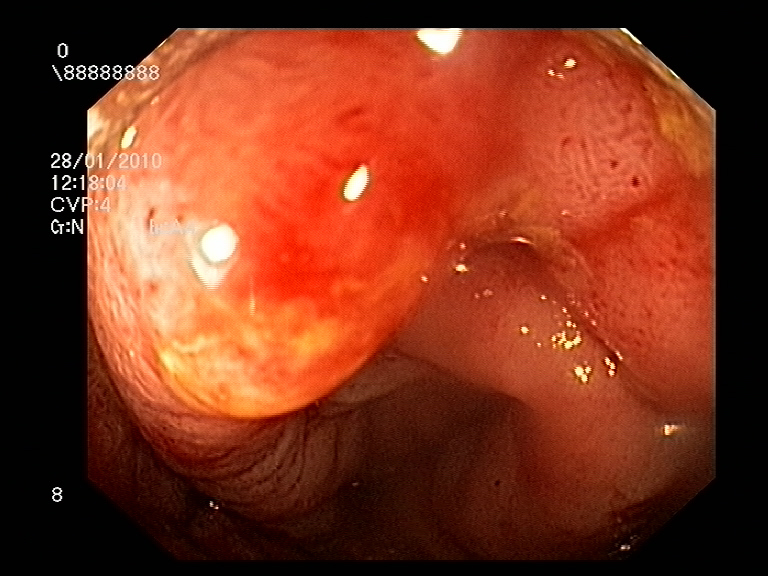}
\label{fig_data_1}}
\hfil
\subfloat[]{\includegraphics[width = 0.24\textwidth,valign=c]{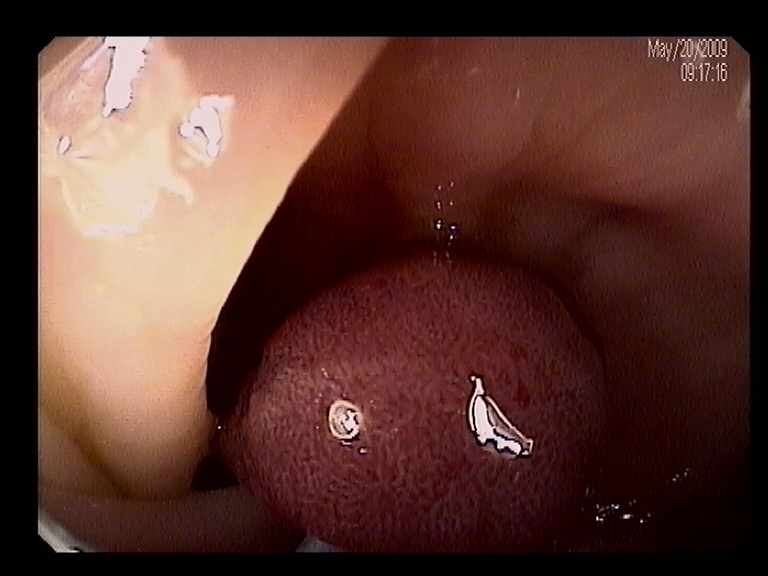}
\label{fig_data_2}}
\hfil
\subfloat[]{\includegraphics[width = 0.24\textwidth,valign=c]{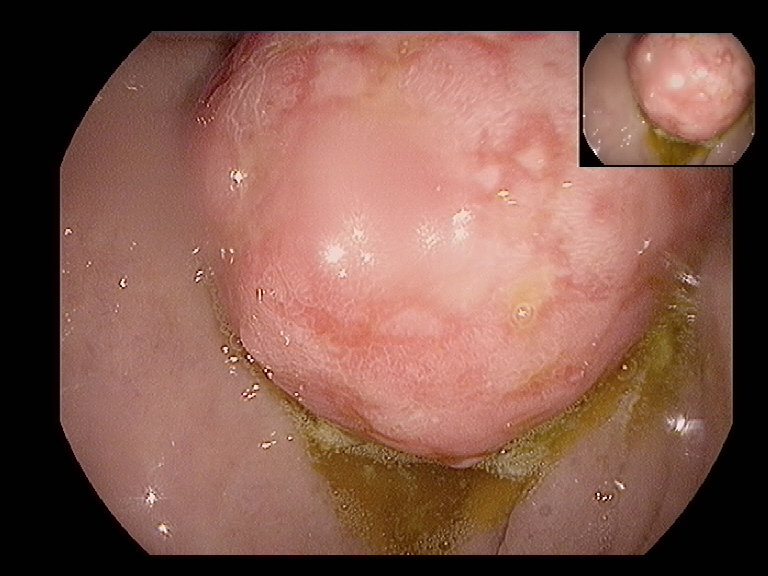}
\label{fig_data_3}}
\hfil
\subfloat[]{\includegraphics[width = 0.24\textwidth,valign=c]{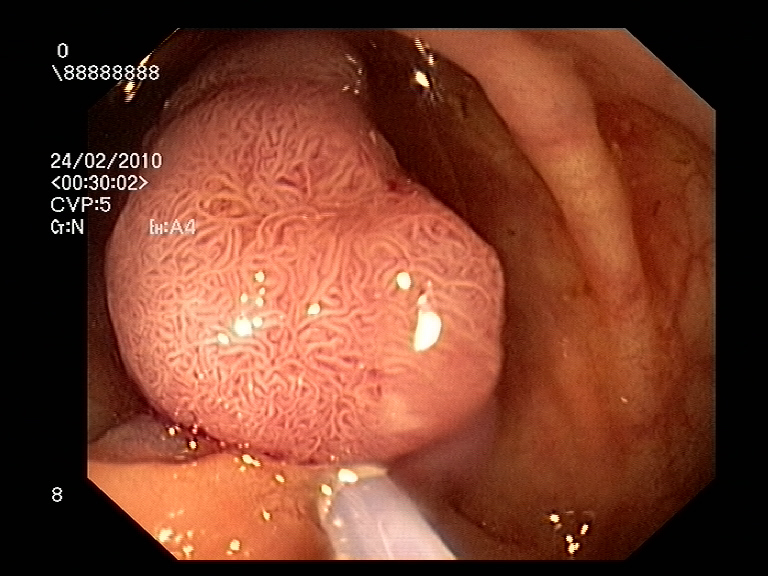}
\label{fig_data_4}}
\caption{Polyp appearance is widely varible in terms of shape, size, and color. Four different polyps extracted from the Kvasir-Seg database \cite{jha_kvasir-seg_2020}.}
\label{fig_datasets}
\end{figure*}

\section{Introduction}
The human gastrointestinal tract can be affected by a variety of pathological abnormalities that may indicate different kind of diseases, ranging from moderately concerning to potentially life-threatening. 
minor annoyances to highly lethal diseases.
For example, Colorectal Cancer (CRC) is the second most common cancer type in women and third most common for men~\cite{haggar_colorectal_2009}. 
Among other pathological findings of interest, gastro-intestinal polyps are known to be early precursors of this class of cancer~\cite{gao_benchmark_2017}, presenting themselves in half of the patients over 50 undergoing screening endoscopies~\cite{sanchez-peralta_deep_2020}. 
This kind of patholgoies show a wide range of shapes and visual appearances, as shown in Fig.~\ref{fig_datasets}, and its identification and segmentation represents a challenging problem for both computational techniques and human specialists.

During colonoscopy screenings, a flexible tube with a light camera mounted on it is inserted into the human body through the rectum to analyze it and look for polyps or other pathologies. 
Early detection of CRC substantially increases survival rates, with screening programs enabling even pre-symptomatic treatment \cite{sanchez-peralta_deep_2020}. 
However, it is estimated that around 6-27\% of polyps are not found during a colonoscopic examination~\cite{ahn_miss_2012}, and it has been recently proven in~\cite{lui_new_2020} that up to 80\% of missed lesions could be detected with effective real-time computer-aided colonoscopic image analysis systems. 
Therefore, computational approaches to endoscopic image analysis have been intensely researched as a tool for enhancing colonoscopic procedures and enhance detection rates, enabling early treatment, and increasing survival rates. 

The most relevant computer-aided tasks associated with in computational endoscopic image analysis are:
\begin{enumerate*}
\item Pathology Detection: Deciding if certain findings/lesions appear in an endoscopic frame~\cite{bernal_comparative_2017}.
\item Pathology Classification: Assigning the frame to one among a range of categories, or predict a pathologies' degree of malignancy ~\cite{carneiro_deep_2020}.
\item Pathology Localization: finding the position (often in terms of a bounding box) of lesions within a frame~\cite{zhang_polyp_2018}. 
\item Pathology Segmentation: delineating the exact lesion contour in a given endoscopic frame~\cite{wickstrom_uncertainty_2020}.  
\end{enumerate*}

In this paper we approach the task of pathology classification and pathology segmentation. 
In the following pages we describe our solution for these two tracks of the 2020 Endotect challenge \cite{hicks_overview_2020} consisting of a 23-class endoscopic image classification and a polyp segmentation task.

\section{Methodology}

\subsection{Endoscopic Image Classification}
In this section we describe our approach to multi-class endoscopic frame classification. We first detail the CNN architecture of choice for this work, and then introduce a hierarchical error function that we minimize for obtaining our solution.
\subsubsection{Convolutional Neural Network - BiT}
When working with deep neural networks for image classification, it is common to employ weights from previously optimized models to favor training and reach higher performances with the need of less annotated data. For RGB images like endoscopic frames, this is a particularly efficient approach, since they share common low-level features with natural imagery. The standard approach in this case is to start the training from weights learned from the Imagenet database, which contains 1.3M images. 
Instead, in this work we leverage the recent work of Big Transfer (BiT, \cite{kolesnikov_big_2020}), and start our training from a model pre-trained on the ImageNet-21k extension, which contains 14M images. 
The architecture of choice in this paper corresponds with ResNet-50x1/BiT-M in \cite{kolesnikov_big_2020}.

\begin{figure*}[t]
\centering
\includegraphics[width = \textwidth]{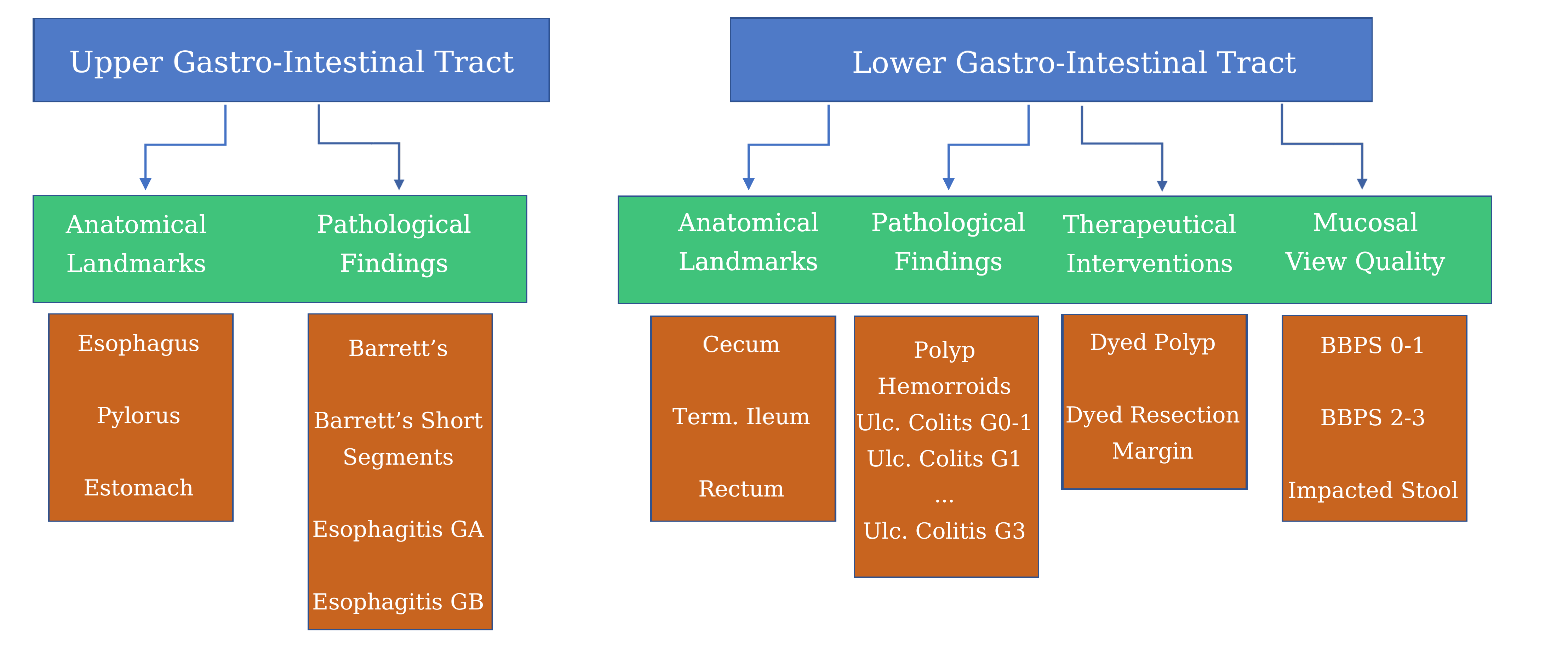}
\label{fig_deg_1}
\caption{Label structure for the endoscopic image classification task of the Endotect Challenge. Blue blocks correspond to Gastro-Intestinal Tract (Lower/Upper), green blocks represent the nature of the finding, and brown blocks the finding. In this work we follow this hierarchy to define a multi-task loss function that minimizes inconsistencies in predictions.}
\label{fig_categories}
\end{figure*}

\subsubsection{Hierarchical Loss Function}
The data used for multi-category classification on endoscopic images for the Endotect challenge was labeled with a rich structure, as illustrated in Fig. \ref{fig_categories}. 
In this work, we leverage such structure by introducing a hierarchical loss function that reflects the different ways in which the same image is labeled. 
For this, an input image $x$ is classified in a three-fold manner: first the network predicts which tract the image belongs to, \textit{e.g.} upper or lower gastro-intestinal tract; second, the category of the finding is also predicted (anatomical landmark, pathological finding, therapeutic intervention, or mucosal view's quality), and last the network outputs the corresponding finding among the 23 different classes. 
The aim of this approach is to minimize inconsistencies, since for example therapeutic interventions only appear in the lower tract, and some pathological findings are only found in the upper tract. 

Each of the aforementioned three classification problems is handled by an independent cross-entropy loss function $\mathcal{L}_{\mathrm{tract}}$, $\mathcal{L}_{\mathrm{cat}}$, $\mathcal{L}_{\mathrm{find}}$; each of these losses attempts to minimize the errors in a hierarchical manner: errors in classifying the tract for a given image will result in a higher error, since it will directly propagate into the category and finding error.

In order to balance the contribution of each of these losses, we consider the amount of classes $n$ and the cross-entropy error that a random prediction would produce in each sub-task, $\log(1/n)$, as follows:
\begin{align}\label{loss}
\mathcal{L}_{\mathrm{overall}}(U(x), y) &= \log(n_{\mathrm{tract}})\cdot \mathcal{L}_{\mathrm{tract}}(U(x), y_{\mathrm{tract}})+\\
 & \log(n_{\mathrm{cat}})\cdot \mathcal{L}_{\mathrm{cat}}(U(x), y_{\mathrm{cat}})+\nonumber\\&\log(n_{\mathrm{find}})\cdot \mathcal{L}_{\mathrm{find}}(U(x), y_{\mathrm{find}})\nonumber
\end{align}
In the above equation, there is a slight abuse of notation: while the network's output is the probability of belonging to each of the 23 different findings, probabilities of higher hierarchical levels (category and tract) are computed by suitably aggregating probabilities of the lower hierarchical levels. 
For example, in order to compute $U(x)$ in the $\mathcal{L}_{\mathrm{tract}}$ case, we would add all the probabilities corresponding to findings that are known to appear only in the lower or upper GI tracts.

\begin{figure*}[t]
\centering
\includegraphics[width = \textwidth]{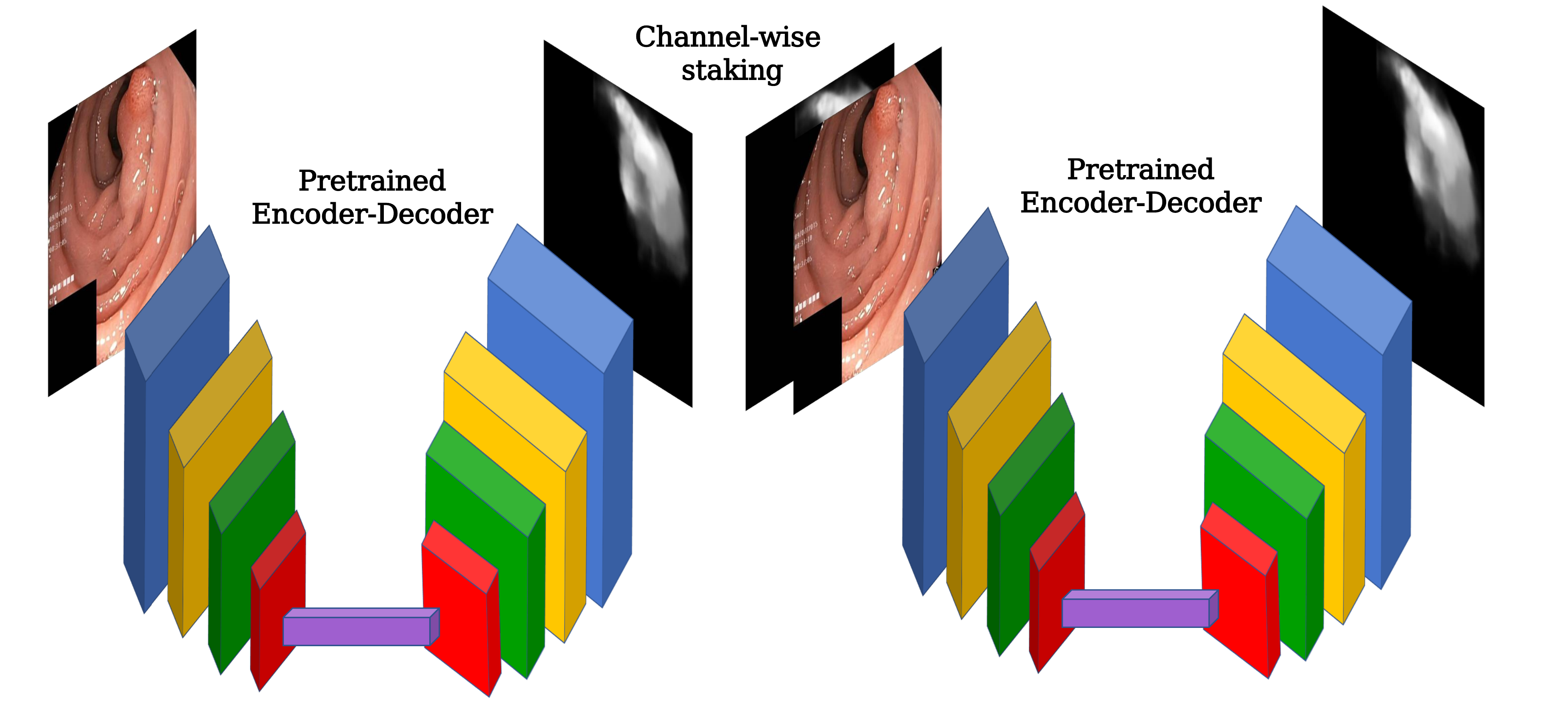}
\label{fig_deg_1}
\caption{Pre-trained Double Encoder-Decoder Network for Polyp Segmentation. The second network receives as input a stack of four channels: the original image together with the prediction of the first network; this facilitates the task of the second network, which can focus on interesting or error-prone areas of the image, enhancing its segmentation capability.}
\label{fig_enc_dec}
\end{figure*}

\subsection{Gastrointestinal Polyp Segmentation}
In the following lines we describe our solution to the polyp segmentation task of the Endotect Challenge.
\subsubsection{Double Autoencoders}
For the segmentation of polyps we adopt the solution presented in \cite{galdran_double_2020}, which is based on a double encoder-decoder network introduced in \cite{galdran_little_2020}. 
Double encoder-encoders are a straightforward extension of encoder-decoder architectures in which two encoder-decoder networks are sequentially combined, as shown in Fig. \ref{fig_enc_dec}.
Being $x$ the input RGB image, $E^{(1)}$ the first CNN, and $E^{(2)}$ the second CNN, a double encoder-decoder feeds the output $E^{(1)}(x)$ of the first network to the second network together with $x$, which effectively acts as an attention map allowing $E^{(2)}$ to focus on the most relevant areas of the image $x$:
\begin{equation}\label{wnet_def}
E(x) = E^{(2)}(x, E^{(1)}(x)),
\end{equation}
where $x$ and $E^{(1)}(x)$ are stacked so that the input to $E^{(2)}$ has four channels instead of the three channels corresponding to the RGB components of $x$. 
Differently from \cite{galdran_double_2020}, we employ a Dual Path Networks DPN92 \cite{chen_dual_2017} as our encoder for both networks, and Feature-Pyramid decoders \cite{lin_feature_2017}.
Supervision is applied at the output of both networks by means of a standard cross-entropy loss.

\subsection{Training Details}
For both classification and segmentation tasks we followed a similar training procedure. 
Both networks were optimized using standard Stochastic Gradient Descent by minimizing the loss in eq. (\ref{loss}) for classification, and the pixel-wise cross-entropy loss for segmentation, with a learning rate of $l=0.01$ and a batch-size of $8$ and $4$ respectively. 
The learning rate was decayed following a cosine law from its initial value to $l=1e-8$ during $n$ epochs, which defines a training cycle. 
In the classification case, $n=10$ whereas in the segmentation case $n=50$. 
We then repeated this process for $5$ cycles in the classification scenario and $20$ cycles in the segmentation case, restarting the learning back at the beginning of each cycle.
Images were re-scaled to $640\times512$, which respects the predominant rectangular aspect ratio of the original data, and were augmented with standard techniques (random rotations, vertical/horizontal flipping, contrast/saturation/brightness changes) during training. 
The mean Dice score is monitored on a separate validation set and the best performing model is kept for testing purposes. 
For testing, we generate four different versions of each image by horizontal/vertical flipping, predict on each of them, and average the results.

\begin{table*}[t]  %
	\renewcommand{\arraystretch}{1.3}	
	\centering
\setlength\tabcolsep{3.5pt}	
\begin{tabular}{l cccccc}
  & \multicolumn{2}{c}{\textbf{Finding Classif.}} & \multicolumn{4}{c}{\textbf{Polyp Segmentation}} \\
\cmidrule(lr){2-3} \cmidrule(lr){4-7}
				                &  MCC &  F1-micro     &  Jaccard &  F1  &  Precision  & Recall       \\
\midrule
\textbf{5-fold CV (Avg.)}       & 91.25  & 91.82        &  88.03   & 92.30 &  93.19 &  94.14      \\
\textbf{Hidden Test Set}        & 85.61  & 86.96        &  87.10   & 91.97 &  92.82 &  93.73     \\
\bottomrule\\[0.05cm]
\end{tabular}
\caption{Performance of our solution to the classification and segmentation tasks. First row: average performance on five-fold cross validation. Second row: results on the hidden test set provided by the organizers.}
\label{tab_results}
\end{table*}%

\section{Experimental Results}
The Hyper-Kvasir dataset, provided in \cite{borgli_hyperkvasir_2020}, was used for training our classification model, whereas the Kvasir-seg dataset~\cite{jha_kvasir-seg_2020} was used for polyp segmentation training purposes. 
In the first case, 10,662 labeled images containing one of the 23 different findings is available, but the frequency of these categories is highly imbalanced. 
We choose to compensate for this by simple oversampling of minority classes. 
In the second case, 1,000 images containing manually delineated polyps are provided by the organization. 
For both the classification and segmentation tasks we carry out a five-fold cross-validation training; the average result is shown in the first row of Table \ref{tab_results}. 
The five models, which are trained on different subsets of the training dataset, are then employed to generate ensembled predictions on two hidden tests set that are submitted to the organization for performance ranking. 
The resulting performance on the test set is also shown in the second row of Table \ref{tab_results}. 
At the time of writing no public leaderboard has been released yet by the organizers.

\section{Discussion and Conclusion}
In this work we have described our solution to the 2020 Endotect challenge, classification and segmentation sub-challenges. 
Our approach to endoscopic image classification relied on a CNN trained so as to minimize in a multi-task manner a hierarchical set of three loss functions that penalize inconsistencies not only in the final finding classification but also in higher levels of classification like the section of the gastrointestinal tract the image comes from or the kind of finding it contains. 
For the segmentation sub-challenge, we leveraged our recent work on polyp segmentation \cite{galdran_double_2020}: a double decoder-encoder network with a DPN92 encoder and a FP-Net decoder was trained to accurately delineate polyp boundaries. 

Although at the time of writing the organizers have not released a public ranking yet, performance in internal cross-validation analysis was promising. Certain overfitting is observed in the classification sub-challenge, but segmentation performance scores on the hidden test set of the challenge seem well-aligned with our average cross-validation indicators.

\bibliographystyle{splncs04}
\bibliography{aiha_polyp.bib}

\begin{thebibliography}{10}
\providecommand{\url}[1]{\texttt{#1}}
\providecommand{\urlprefix}{URL }
\providecommand{\doi}[1]{https://doi.org/#1}

\bibitem{ahn_miss_2012}
Ahn, S.B., Han, D.S., Bae, J.H., Byun, T.J., Kim, J.P., Eun, C.S.: The {Miss}
  {Rate} for {Colorectal} {Adenoma} {Determined} by {Quality}-{Adjusted},
  {Back}-to-{Back} {Colonoscopies}. Gut and Liver  \textbf{6}(1),  64--70 (Jan
  2012)

\bibitem{bernal_comparative_2017}
Bernal, J., Tajkbaksh, N., Sánchez, F.J., Matuszewski, B.J., Chen, H., Yu, L.,
  Angermann, Q., Romain, O., Rustad, B., Balasingham, I., Pogorelov, K., Choi,
  S., Debard, Q., Maier-Hein, L., Speidel, S., Stoyanov, D., Brandao, P.,
  Córdova, H., Sánchez-Montes, C., Gurudu, S.R., Fernández-Esparrach, G.,
  Dray, X., Liang, J., Histace, A.: Comparative {Validation} of {Polyp}
  {Detection} {Methods} in {Video} {Colonoscopy}: {Results} {From} the {MICCAI}
  2015 {Endoscopic} {Vision} {Challenge}. IEEE Transactions on Medical Imaging
  \textbf{36}(6),  1231--1249 (Jun 2017)

\bibitem{borgli_hyperkvasir_2020}
Borgli, H., Thambawita, V., Smedsrud, P.H., Hicks, S., Jha, D., Eskeland, S.L.,
  Randel, K.R., Pogorelov, K., Lux, M., Nguyen, D.T.D., Johansen, D., Griwodz,
  C., Stensland, H.K., Garcia-Ceja, E., Schmidt, P.T., Hammer, H.L., Riegler,
  M.A., Halvorsen, P., de~Lange, T.: {HyperKvasir}, a comprehensive multi-class
  image and video dataset for gastrointestinal endoscopy. Scientific Data
  \textbf{7}(1), ~283 (2020)

\bibitem{carneiro_deep_2020}
Carneiro, G., Zorron Cheng Tao~Pu, L., Singh, R., Burt, A.: Deep learning
  uncertainty and confidence calibration for the five-class polyp
  classification from colonoscopy. Medical Image Analysis  \textbf{62},  101653
  (May 2020)

\bibitem{chen_dual_2017}
Chen, Y., Li, J., Xiao, H., Jin, X., Yan, S., Feng, J.: Dual path networks. In:
  Proceedings of the 31st {International} {Conference} on {Neural}
  {Information} {Processing} {Systems}. pp. 4470--4478. {NIPS}'17, Curran
  Associates Inc., Red Hook, NY, USA (Dec 2017)

\bibitem{galdran_little_2020}
Galdran, A., Anjos, A., Dolz, J., Chakor, H., Lombaert, H., Ayed, I.B.: The
  {Little} {W}-{Net} {That} {Could}: {State}-of-the-{Art} {Retinal} {Vessel}
  {Segmentation} with {Minimalistic} {Models}. arXiv:2009.01907  (Sep 2020)

\bibitem{galdran_double_2020}
Galdran, A., González~Ballester, M.A., Carneiro, G.: Double
  {Encoder}-{Decoder} {Networks} for {Gastrointestinal} {Polyp} {Segmentation}.
  In: {ICPR} {Workshop} on {Artificial} {Intelligence} for {Healthcare}
  {Applications} (2020)

\bibitem{haggar_colorectal_2009}
Haggar, F.A., Boushey, R.P.: Colorectal {Cancer} {Epidemiology}: {Incidence},
  {Mortality}, {Survival}, and {Risk} {Factors}. Clinics in Colon and Rectal
  Surgery  \textbf{22}(4),  191--197 (Nov 2009)

\bibitem{hicks_overview_2020}
Hicks, S., Jha, D., Thambawita, V., Halvorsen, P., Hammer, H., Riegler, M.: An
  {Overview} of the {EndoTect} {Challenge} at {ICPR} 2020. In: Proceedings in
  the 25th {International} {Conference} on {Pattern} {Recognition} ({ICPR})
  (2020)

\bibitem{jha_kvasir-seg_2020}
Jha, D., Smedsrud, P.H., Riegler, M.A., Halvorsen, P., de~Lange, T., Johansen,
  D., Johansen, H.D.: Kvasir-seg: {A} segmented polyp dataset. In:
  International {Conference} on {Multimedia} {Modeling}. pp. 451--462. Springer
  (2020)

\bibitem{kolesnikov_big_2020}
Kolesnikov, A., Beyer, L., Zhai, X., Puigcerver, J., Yung, J., Gelly, S.,
  Houlsby, N.: Big {Transfer} ({BiT}): {General} {Visual} {Representation}
  {Learning}. In: Vedaldi, A., Bischof, H., Brox, T., Frahm, J.M. (eds.)
  Computer {Vision} – {ECCV} 2020. pp. 491--507. Lecture {Notes} in
  {Computer} {Science}, Springer International Publishing, Cham (2020)

\bibitem{lin_feature_2017}
Lin, T.Y., Dollár, P., Girshick, R., He, K., Hariharan, B., Belongie, S.:
  Feature {Pyramid} {Networks} for {Object} {Detection}. In: 2017 {IEEE}
  {Conference} on {Computer} {Vision} and {Pattern} {Recognition} ({CVPR}). pp.
  936--944 (Jul 2017), iSSN: 1063-6919

\bibitem{lui_new_2020}
Lui, T.K., Hui, C.K., Tsui, V.W., Cheung, K.S., Ko, M.K., aCC Foo, D., Mak,
  L.Y., Yeung, C.K., Lui, T.H., Wong, S.Y., Leung, W.K.: New insights on missed
  colonic lesions during colonoscopy through artificial intelligence–assisted
  real-time detection (with video). Gastrointestinal Endoscopy  (May 2020)

\bibitem{sanchez-peralta_deep_2020}
Sánchez-Peralta, L.F., Bote-Curiel, L., Picón, A., Sánchez-Margallo, F.M.,
  Pagador, J.B.: Deep learning to find colorectal polyps in colonoscopy: {A}
  systematic literature review. Artificial Intelligence in Medicine
  \textbf{108},  101923 (Aug 2020)

\bibitem{gao_benchmark_2017}
Vázquez, D., Bernal, J., Sánchez, F.J., Fernández-Esparrach, G., López,
  A.M., Romero, A., Drozdzal, M., Courville, A.: A {Benchmark} for
  {Endoluminal} {Scene} {Segmentation} of {Colonoscopy} {Images}. Journal of
  Healthcare Engineering  \textbf{2017},  4037190 (Jul 2017)

\bibitem{wickstrom_uncertainty_2020}
Wickstrøm, K., Kampffmeyer, M., Jenssen, R.: Uncertainty and interpretability
  in convolutional neural networks for semantic segmentation of colorectal
  polyps. Medical Image Analysis  \textbf{60},  101619 (Feb 2020)

\bibitem{zhang_polyp_2018}
Zhang, R., Zheng, Y., Poon, C.C.Y., Shen, D., Lau, J.Y.W.: Polyp detection
  during colonoscopy using a regression-based convolutional neural network with
  a tracker. Pattern Recognition  \textbf{83},  209--219 (Nov 2018)

\end{thebibliography}

\end{document}